%% file: PtyNISQ.tex
\documentclass[%
 aip,
 groupedaddress, 
 amsmath,amssymb,
 reprint,%
]{revtex4-1}

\usepackage{graphicx}
\usepackage{dcolumn}
\usepackage{bm}

\usepackage[utf8]{inputenc}
\usepackage[T1]{fontenc}
\usepackage{mathptmx}
\usepackage{etoolbox}

\usepackage{natbib}
\usepackage{amsmath}
\usepackage{epsfig}
\usepackage{amssymb}
\usepackage{hyperref}
\hypersetup{colorlinks=true,linkcolor=blue,citecolor=blue,urlcolor=blue}
\usepackage{times}
\usepackage{algorithm}
\usepackage[noend]{algpseudocode}
\usepackage{qcircuit}

\usepackage{xfrac}    
\usepackage{nicefrac} 

\makeatletter
\def\@email#1#2{%
 \endgroup
 \patchcmd{\titleblock@produce}
  {\frontmatter@RRAPformat}
  {\frontmatter@RRAPformat{\produce@RRAP{*#1\href{mailto:#2}{#2}}}\frontmatter@RRAPformat}
  {}{}
}%
\makeatother
\begin{document}

\preprint{AIP/123-QED}

\title{\huge Ptychographic estimation of pure multiqubit states in a quantum device \vspace{3mm}}

\author{Warley M. S. Alves}
\author{Leonardo Neves\vspace{2mm}}%
\altaffiliation[Corresponding author: ]{lneves@fisica.ufmg.br}
\affiliation{\normalsize\text{Departamento de F\'isica, Universidade Federal de Minas Gerais, Belo Horizonte, MG, Brazil}}%


\begin{abstract}
Quantum ptychography is a method for estimating an unknown pure quantum state by subjecting it to overlapping projections, each one followed by a projective measurement on a single prescribed basis. Here, we present a comprehensive study of this method applied for estimating $n$-qubit states in a circuit-based quantum computer, including numerical simulations and experiments carried out on an IBM superconducting quantum processor. The intermediate projections are implemented through Pauli measurements on one qubit at a time, which sets the number of ptychographic circuits to $3n$ (in contrast to the $3^n$ circuits for standard Pauli tomography); the final projective measurement in the computational basis is preceded by the quantum Fourier transform (QFT). Due to the large depth and number of two-qubit gates of the QFT circuit, which is unsuitable for noisy devices, we also test the approximate QFT (AQFT) and separable unitary operations. Using the QFT and AQFT of degree $2$, we obtained high estimation fidelities in all tests with separable and entangled states for up to three and four qubits, respectively; on the other hand, the separable unitaries in this scenario provided good estimations only for separable states, in general. Our results compare favorably with recent results in the literature and we discuss further alternatives to make the ptychographic method scalable for the current noisy devices. 
\end{abstract}

\maketitle

\section{Introduction}

Quantum computing offers a new approach to computational tasks in which quantum effects are exploited to process information in ways that are impossible for classical machines. In this way, challenging problems in cryptography, optimization, and simulation that are intractable for a classical computer, could be solved efficiently in its quantum counterpart.\cite{Bennett2000,NielsenBook} Recently, we have witnessed the emergence of intermediate-scale quantum processors (i.e., devices with at most a few hundred qubits) based on different physical platforms, such as ion traps,\cite{Schindler2013,Bruzewicz2019,Pino2021,Moses2023} neutral atoms,\cite{Saffman2016,Henriet2020,Barnes2022} photonic systems,\cite{Takeda2019,Slussarenko2019,Arrazola2021} and superconducting circuits,\cite{Clarke2008,Gambetta2017,Kjaergaard2020} among others. Due to their limited number of qubits, and high susceptibility to noise and decoherence, the state-of-the-art processors have been described as noisy intermediate-scale quantum devices.\cite{Preskill2018} While these devices do not seem able to fulfill the promising achievements of a quantum computer in the short term, they are already being used, for instance, to taste what would be the so-called \emph{quantum computational advantage}\cite{Arute2019,Zhong2020} and perform small simulations of physical and chemical systems.\cite{Ippoliti2021,Eddins2022,Motta2023}

The development of methods to evaluate the performance of quantum processors is a relevant task not only for current devices but also for envisioning an era of fault-tolerant quantum computing. Methods such as randomized benchmarking to measure the average error rates of quantum gates\cite{Magesan2012,Helsen2022} or direct fidelity estimation of quantum states from a few specific measurements\cite{Silva2011,Flammia2011} are important in this context. However, they are not as complete as quantum state estimation,\cite{Banaszek2013,TeoBook} which allows us to acquire full information about the state of a quantum system and thus assess the performance of a device, in principle, at any stage of the computation. Unfortunately, this comes at the high cost of requiring a number of measurements that increases exponentially with the number of qubits,\cite{James2001} $n$, as well as a computational postprocessing that may become impractical for large $n$.\cite{Shang2017} Several methods have been suggested to overcome these issues by exploiting prior information about the state to be determined. For example, prior knowledge that a quantum state has a low rank or is close to being permutationally invariant allows it to be determined with a significant reduction in both the number of measurements and postprocessing times through compressed sensing\cite{Gross2010} or permutationally invariant estimation\cite{Toth2010} methods, respectively. Furthermore, considerable efforts have been devoted to developing methods for estimating pure states,\cite{Ferrie2014,Goyeneche2015,Ma2016,Stefano2017,Fernandes2019,Zambrano2020,Pereira2022,Zambrano2024,Tariq2024} highlighting the topical relevance of the subject. Even though purity is at best a good approximation in realistic scenarios, the state estimated in these cases is expected to be close to the actual one and is obtained with much fewer resources than standard estimation schemes. 

One such method for pure state estimation, called quantum ptychography, was proposed by Fernandes and Neves\cite{Fernandes2019} as an analogue of the powerful computational imaging technique used in optical\cite{Thibault2008} and electron\cite{Humphry2012} microscopy. In classical ptychography,\cite{Faulkner04,Rodenburg04} the complex-valued transmission function of an object is computationally estimated by an iterative algorithm that processes multiple diffraction patterns generated by shifting a coherent illumination over partially overlapping regions of the object. In its quantum counterpart, an unknown pure state, characterized by its complex amplitudes, is the object to be determined; the role of the shifting illuminations and the corresponding diffraction patterns is played by projections onto partially overlapping subspaces of the state space and a projective measurement in a single prescribed basis, respectively. The data generated in this process are then fed into an iterative phase retrieval algorithm that estimates the state. The effectiveness of quantum ptychography relies on a sufficient diversity and redundancy in the data, determined by the overlapping projections. If achieved, the algorithm will make an initial random guess converge to the true pure state that generated the data. This is done by using the measurement outcomes to iteratively impose corrections to multiple overlapping ``slices'' of the estimate and updating it until convergence.

In this work, we present a comprehensive study of the ptychographic method applied for estimating $n$-qubit pure states in a circuit-based quantum computer. Our study comprises numerical simulations as well as experiments performed on an IBM superconducting quantum processor. Following the original proposal,\cite{Fernandes2019} the overlapping projections are implemented through Pauli measurements on one qubit at a time, followed by a quantum Fourier transform (QFT) and a projective measurement in the computational basis. Since the virtual quantum circuit for an $n$-qubit QFT consists of $n$ single-qubit and $O(n^2)$ two-qubit gates,\cite{Coppersmith1994,NielsenBook} its implementation in a real quantum processor with limited connectivity and native operations will generate circuits with even more gates and larger depths, making them highly susceptible to noise and decoherence. As an alternative to mitigate these effects, we propose and test the use of the approximate QFT (AQFT) in the ptychographic method.  This operation, introduced by Coppersmith,\cite{Coppersmith1994} is parametrized by an integer $m$ ($1\leqslant m\leqslant n$) that sets the degree of approximation (e.g., $m=n$ corresponds to the exact QFT, while $m=1$ results in the Hadamard transform). The smaller the value of $m$, the farther the approximation is from the exact QFT. However, this also reduces the number of two-qubit gates, thereby generating less noisy and lower depth circuits in a real processor.  Using the exact QFT and AQFT with a degree of $m=2$, we obtained high experimental estimation fidelities in all tests with separable and entangled states for up to $n=3$ and $n=4$ qubits, respectively, attesting to the better performance of the latter operation. In addition to QFT and AQFT, we also implemented the ptychographic method with separable unitary operations, which generate less noisy circuits as they do not require two-qubit gates. However, in this case, we obtained good estimations only for separable states, in general. The experimental fidelities were achieved by employing a simple measurement error mitigation technique and introducing a modification to the original estimation algorithm, which improved its convergence. 

To evaluate performance and discuss the advantages and disadvantages of the ptychographic method from the perspectives of estimation accuracy and both experimental and computational costs, we compare it with two other techniques. The former is the well-known standard tomographic scheme based on Pauli measurements \cite{James2001,Altepeter2005}, designed for arbitrary states. The second one is recently introduced by Pereira, Zambrano, and Delgado (PZD) \cite{Pereira2022}, and is also designed for pure multiqubits states. For both ptychographic and PZD methods, the number of circuits increases linearly with $n$, while in the Pauli tomography, this increasing is exponential. Quantum ptychography, on the other hand, has the disadvantage of requiring more complex circuits due to the QFT (or even the AQFT), while Pauli and PZD methods are performed only with single-qubit gates. Despite this, we will show that our results, from both simulations and experiments, compared favorably with those from PZD, demonstrating the effectiveness and potential of the ptychographic method for estimating multiqubit states and becoming a valuable tool for performance assessment in a quantum computer. 

This article is organized as follows: in Sec.~\ref{sec:PtyQ}, we describe the multiqubit ptychographic method and its implementation in a circuit-based quantum computer. In Sec.~\ref{sec:Prelim}, we outline the operational aspects of the IBM quantum processor and a simple error mitigation technique employed in our experiments. Section~\ref{sec:Results} is devoted to presenting our results from both numerical simulations and experiments using the QFT in the protocol. In Sec.~\ref{sec:ExperimentAQFT}, we study the ptychographic method under different unitary operations and present experimental results obtained with the AQFT and separable unitaries. In Sec.~\ref{sec:Discussion}, we analyze our experimental results and compare them with those from Ref.~\onlinecite{Pereira2022};  in addition, we discuss further steps to make quantum ptychography scalable for noisy devices. Finally, in Sec.~\ref{sec:Conclusion}, we present our concluding remarks.

\section{Quantum ptychography for multiqubit pure states}
\label{sec:PtyQ}

\subsection{Overview of the protocol}
\label{subsec:Ptycho_overview}
A pure state of an $n$-qubit system can be written in the computational basis as
\begin{align}   \label{eq:Psi}
|\psi\rangle=\sum_{j=0}^{2^n-1}\alpha_j|j\rangle,
\end{align}
where $\sum_j|\alpha_j|^2=1$ and $j=0,\ldots,2^n-1$ is the decimal representation of the $n$-bit binary string $j_{n-1}\cdots j_1j_0$ ($j_k\in\{0,1\}$ $\forall k$). In the ptychographic method,\cite{Fernandes2019} the complex amplitudes $\{\alpha_j\}$, which fully characterize the state, are estimated using the following ingredients:  (i) a set of $N$ projectors $\{\bm{\Pi}_\ell\}_{\ell=0}^{N-1}$ of rank $r$ ($1<r<2^n$), which address all levels of the $2^n$-dimensional Hilbert space, $\mathcal{H}_n$, and satisfy the partial overlapping condition $0<\textrm{Tr}(\bm{\Pi}_{\ell}\bm{\Pi}_{{\ell}^{\prime}})/r<1$ for at least one partner; (ii) a projective measurement in the basis $\{\mathbf{U}^\dag|j\rangle\}_{j=0}^{2^n-1}$, where $\mathbf{U}$ is a unitary operator acting on $\mathcal{H}_n$. With a suitable choice of $\{\bm{\Pi}_\ell\}$ and $\mathbf{U}$, the protocol proceeds as follows: first, one applies a $\bm{\Pi}_\ell$ on the input ensemble described by the state $|\psi\rangle$ and then the prescribed projective measurement is performed on the output described by $|\psi_\ell\rangle=\bm{\Pi}_\ell|\psi\rangle$, generating the count distribution $\Omega_\ell=\{|\langle j|\mathbf{U}|\psi_\ell\rangle|^2\}_{j=0}^{2^n-1}$. This procedure is repeated for each projector $\bm{\Pi}_\ell$ and gives us the dataset $\{\sqrt{\Omega_\ell}\}_{\ell=0}^{N-1}$ that will feed an iterative phase retrieval algorithm designed to estimate the state. The standard algorithm for this task is known as ptychographic iterative engine (PIE) \cite{Faulkner04,Rodenburg04} and can be visualized by the pseudocode in Algorithm~\ref{alg:PIE}; further details will be provided in Sec.~\ref{subsec:PIE}. 

\begin{algorithm}[H]
	\caption{PIE algorithm.}
	\begin{algorithmic}
	\State{Define $|\varphi_0\rangle$} \Comment{Input random estimate}
    \For{$j=1,\ldots,N_{\textsc{pie}}$} \Comment{Loop over PIE iterations}
	\State $|\varphi_j\rangle \gets |\varphi_{j-1}\rangle$ 
    \For{$\ell=0,\ldots,N-1$} \Comment{Loop over projections}
		\State $|\varphi_{j,\ell}\rangle \gets \bm{\Pi}_\ell|\varphi_j\rangle$
		\State $|\tilde{\varphi}_{j,\ell}\rangle \gets \mathbf{U}|\varphi_{j,\ell}\rangle$
		\State $|\tilde{\varphi}_{j,\ell}^{\textrm{corr}}\rangle \gets \sqrt{\Omega_\ell}$ \Comment{Correct amplitudes with data}
		\State $|\varphi_{j,\ell}^{\textrm{corr}}\rangle \gets \mathbf{U}^\dag|\tilde{\varphi}_{j,\ell}^{\textrm{corr}}\rangle$ 
		\State $|\varphi_{j}\rangle \gets |\varphi_{j}\rangle + \beta\bm{\Pi}_\ell(|\varphi_{j,\ell}^{\textrm{corr}}\rangle - |\varphi_{j,\ell}\rangle)$ \Comment{Update estimate}
    \EndFor
	\State $\beta \gets \beta-\Delta\beta$ 
\Comment{Feedback parameter}
	\State Compute figures of merit regarding $|\varphi_{j}\rangle$ 
    \If{(stopping criterion is met)}
		\State \textbf{return} $|\varphi\rangle$
    \EndIf
		\algstore*{Insert}
	\end{algorithmic}
	\label{alg:PIE}
\end{algorithm}

In Ref.~\onlinecite{Fernandes2019}, the authors show that a suitable choice for $\mathbf{U}$ is the QFT 
\begin{align}    \label{eq:QFT}
\mathbf{U}\rightarrow\mathbf{U}_\textsc{qft}=\frac{1}{\sqrt{2^n}}\sum_{j,k=0}^{2^n-1}e^{2\pi i jk/2^n}|j\rangle\langle k|,
\end{align}
while the intermediate overlapping projections are
\begin{align}   \label{eq:Pi_xiq}
\bm{\Pi}_\ell\rightarrow\bm{\Pi}_{\xi q}^\pm=\bm{\pi}_{\xi q}^\pm\otimes\mathbf{I}^{\otimes n-1},
\end{align}
where $\bm{\pi}_{\xi}^\pm=|\xi^\pm\rangle\langle\xi^\pm|$, for $\xi=x,y,z$, is the projector associated with the eigenvalue $\pm 1$ of the corresponding Pauli operator ($\mathbf{X},\mathbf{Y},\mathbf{Z}$), acting on qubit $q=0,\ldots,n-1$; $\mathbf{I}$ denotes the single qubit identity. In addition to satisfying the conditions described above, these projectors ensure sufficient diversity and redundancy in the data, which are essential for the method to succeed.

\subsection{Implementation in a circuit-based quantum computer}

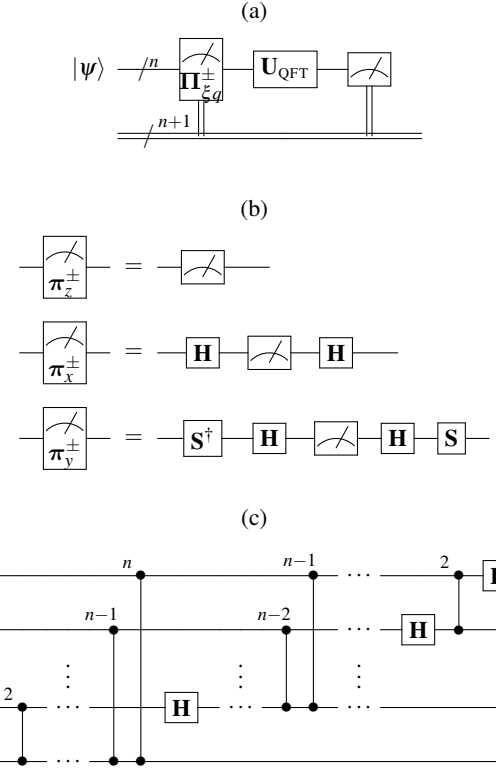
\begin{figure}[t]
\centerline{
\Qcircuit @C=1em @R=1em {
\textrm{(a)} \\ \\
}}
\centerline{
\Qcircuit @C=1.3em @R=1.5em {
 &\lstick{|\psi\rangle}&  {/^n} \qw & \meterB{\bm{\Pi}_{\xi q}^\pm} & \gate{\mathbf{U}_\textsc{qft}} & \meter  \\
 && {/} \cw 	& \cw_{n+1}\cwx[-1] & \cw & \cw\cwx[-1] & \cw \\ \\ \\
}}
\centerline{
\Qcircuit @C=1em @R=1em {
\textrm{(b)} \\ \\
}}
    \centerline{
      \Qcircuit @C=1em @R=1em { 
 & \meterB{\bm{\pi}_z^\pm} & \qw & = && \meter  & \qw  
\\
 & \meterB{\bm{\pi}_x^\pm} & \qw & = && \gate{\mathbf{H}} & \meter & \gate{\mathbf{H}} & \qw  
\\
 & \meterB{\bm{\pi}_y^\pm} & \qw & = && \gate{\mathbf{S}^\dag} & \gate{\mathbf{H}} & \meter & \gate{\mathbf{H}} & \gate{\mathbf{S}} & \qw  
\\ \\ \\
    }}
\centerline{
\Qcircuit @C=1em @R=1em {
\textrm{(c)} \\ \\
}}
\centerline{
  \Qcircuit @C=.9em @R=1em {
& \qw & \qw & \qw & \qw & \qw & \qw & \ctrl{4}_{n} & \qw & \qw & \qw & \qw & \qw  & \ctrl{3}_{n-1}  &  \qw & \cdots &&  \qw & \ctrl{1}_{\;\;\;2} & \gate{\mathbf{H}} & \qw  
\\
& \qw & \qw & \qw & \qw & \qw & \ctrl{3}_{n-1}  & \qw & \qw & \qw & \qw & \qw  & \ctrl{2}_{n-2}  &  \qw & \qw &  \cdots &&  \gate{\mathbf{H}}  & \control \qw & \qw  & \qw  
\\
 &&&&\vdots & & & & & &  \vdots & & & & & \vdots  &&&&&& 
\\   
& \qw  & \ctrl{1}_{\;\;\;2} & \qw & \cdots && \qw & \qw & \gate{\mathbf{H}} & \qw & \cdots &&  \control  \qw & \control\qw & \qw & \cdots && \qw & \qw & \qw & \qw  
\\
& \gate{\mathbf{H}} & \control \qw & \qw & \cdots && \control\qw  & \control \qw & \qw & \qw & \qw & \qw & \qw & \qw & \qw & \qw & \qw & \qw & \qw & \qw & \qw  
\\ 
 } 
}
\caption{(a) An $n$-qubit ptychographic circuit using the overlapping projections given by Eq.~(\ref{eq:Pi_xiq}) and the QFT [Eq.~(\ref{eq:QFT})]. (b) Circuits for the single-qubit Pauli measurements that project the input state in their corresponding eigenvectors $|\xi^\pm\rangle$ ($\xi=x,y,z$); $\mathbf{H}$ and $\mathbf{S}$ denote the Hadamard and phase gates, respectively. (c) $n$-qubit QFT circuit without swap gates. The line segment connecting two solid dots with a superscript $2\leqslant\lambda\leqslant n$ represents the controlled-phase gate $\mathbf{CP}^\lambda_{ij}$ between qubits $i$ and $j$. This representation highlights the symmetry of the gate ($\mathbf{CP}^\lambda_{ij}=\mathbf{CP}^\lambda_{ji}$), which applies a phase-shift $2\pi/2^\lambda$ when both the target and control qubits are in the $|1\rangle$ state.\cite{NielsenBook}} 
\label{fig:Pty_circuit} 
\end{figure}

In a circuit-based quantum computer, the $N=6n$ projectors of rank $r=2^{n-1}$ given by Eq.~(\ref{eq:Pi_xiq}) are implemented through Pauli measurements on one qubit at a time, which sets the number of ptychographic circuits to $3n$. Figure~\href{fig:Pty_circuit}{\ref{fig:Pty_circuit}(a)} shows an $n$-qubit ptychographic circuit for this choice of $\{\bm{\Pi}_\ell\}$ and $\mathbf{U}$. The classical register stores $n+1$ bits, one for the intermediate measurement and $n$ for the final one.

The $n$-qubit states resulting from the intermediate projections (\ref{eq:Pi_xiq}) can be written as
\begin{align}    \label{eq:psi_xiq}
\bm{\Pi}_{\xi q}^\pm|\psi\rangle=|\psi_{\xi q}^\pm\rangle=|\xi^\pm\rangle_q\otimes|\phi_{\xi q}^\pm\rangle_{n-1}
\end{align}
and are crucial to the sequence of the protocol. Thus, it is necessary to implement Pauli measurements that project the state of the $q$th qubit into their corresponding eigenstates. In a quantum device where it is only possible to measure in the computational basis, the quantum circuits to accomplish this are shown in Fig.~\href{fig:Pty_circuit}{\ref{fig:Pty_circuit}(b)}; $\mathbf{H}$ and $\mathbf{S}$ are the Hadamard and phase gates, respectively.

To conclude the protocol, one applies the QFT on the states $\{|\psi^\pm_{\xi q}\rangle\}$ of Eq.~(\ref{eq:psi_xiq}) and performs a projective measurement in the computational basis. The $n$-qubit QFT circuit is shown in Fig.~\href{fig:Pty_circuit}{\ref{fig:Pty_circuit}(c)}. In this case, the swap gates at the end of the circuit are not necessary because their role can be accomplished by rearranging the measurement outcomes in the classical register in a way that the output bit string is correct. This also reduces the number of \textsc{cnot} gates in the circuit ($\textsc{\textbf{swap}}_{ij}=\textsc{\textbf{cnot}}_{ij}\textsc{\textbf{cnot}}_{ji}\textsc{\textbf{cnot}}_{ij}$) and, consequently, noise effects.

\subsection{Quantum state estimation based on Pauli measurements and the PZD method}   \label{subsec:PauliPZD}

Throughout this work, we will compare the ptychographic method outlined above with the standard Pauli tomography\cite{James2001,Altepeter2005} and PZD\cite{Pereira2022} methods for estimating unknown multiqubit states, evaluating them in terms of experimental cost, computational postprocessing, and the results obtained in simulations and experiments carried out in equivalent scenarios. The former is based on Pauli measurements and designed for arbitrary states: an arbitrary $n$-qubit density matrix can be written in the operator basis $\{\mathbf{I},\mathbf{X},\mathbf{Y},\mathbf{Z}\}^{\otimes n}\equiv\{\bm{\Gamma}_j\}_{j=1}^{4^n}$ as $\hat{\rho}=\frac{1}{2^n}\sum_{j=1}^{4^n}c_j\bm{\Gamma}_j$, where $c_j=\textrm{Tr}(\hat{\rho}\bm{\Gamma}_j)$. To determine this state, it is sufficient to perform projective measurements in the $3^n$ bases whose operator $\bm{\Gamma}_j$ does not include any $\mathbf{I}$ in the tensor product.\footnote{The probabilities that build $c_j$ from operators including $\mathbf{I}$ can be determined from the measurements in the other $3^n$ bases.} Similar to quantum ptychography, the second method, PZD, is designed for pure states. It requires projective measurements in $\mu n+1$ separable bases ($\mu\geqslant 2$): the measurement in the computational basis determines the magnitudes $\{|\alpha_j|\}$ of the coefficients in Eq.~(\ref{eq:Psi}) whereas the measurements in the other $\mu n$ bases determine their phases $\{\arg\alpha_j\}$ by solving a sequence of systems of linear equations. This algorithm succeeds by a suitable choice of $\mu$ and the measurement bases.\footnote{The PZD method can also be implemented from measurements on $\mu$ nonseparable bases plus the computational basis. However, this approach is less favorable for noisy devices and achieved poor performance in comparison to the use of separable bases.\cite{Pereira2022}}

In both ptychographic and PZD methods, the number of circuits increases linearly with the number of qubits ($3n$ and $\mu n+1$, respectively), in contrast to the exponential increase of the Pauli tomography ($3^n$). Obviously this occurs because the first methods are restricted to pure states, which are characterized by a considerably smaller number of real parameters than a density matrix. On the other hand, both Pauli and PZD methods are performed in circuits with short and nonincreasing depth that contain only single-qubit gates. In contrast, the ptychographic circuits [Fig.~\href{fig:Pty_circuit}{\ref{fig:Pty_circuit}(a)}] are more complex, especially due to the QFT whose circuit [Fig.~\href{fig:Pty_circuit}{\ref{fig:Pty_circuit}(c)}] has a depth of $2n$ and contains $n$ single-qubit gates and $n(n-1)/2$ two-qubit gates. By this criterion, the ptychographic method is less suitable in the current scenario of noisy devices. However, we will see that there are alternatives to get around this problem, for instance, using other unitary operations instead of the QFT.

\section{Preliminaries}
\label{sec:Prelim}

When executing the built circuits in a quantum processor, it is important to know some basic operational aspects that will be useful to understand its limitations, interpret the results obtained, and, possibly, optimize the algorithms. Here, we address issues such as the limited set of native gates and connectivity and describe the general procedures adopted for the experiments of quantum ptychography, which were carried out at ibm\_perth, a seven-qubit superconducting device whose main features are summarized in Table~\ref{tab:ibm_devices}.

\begin{table}[b]
\caption{Main features of the IBM device used in this work.}
\label{tab:ibm_devices} 
\begin{ruledtabular}
\begin{tabular}{ll}
Hardware &  ibm\_perth \\
Number of qubits & 7 \\
Native gates & $\mathbf{I},\mathbf{X},\mathbf{\sqrt{X}},\mathbf{R}_z(\theta),\textsc{\textbf{cnot}}$ \\
Coupling map & \textsf{H}-like \\
Average errors of local gates & $2.5\times 10^{-4}$ \\
Average errors of \textsc{cnot} gates & $8\times 10^{-3}$ \\
Average readout errors & $2.5\times 10^{-2}$ \\
Number of shots & $10^5$ \\
\end{tabular}
\end{ruledtabular}
\end{table}

\begin{figure*}[t]
\centering
\includegraphics[width=1\textwidth]{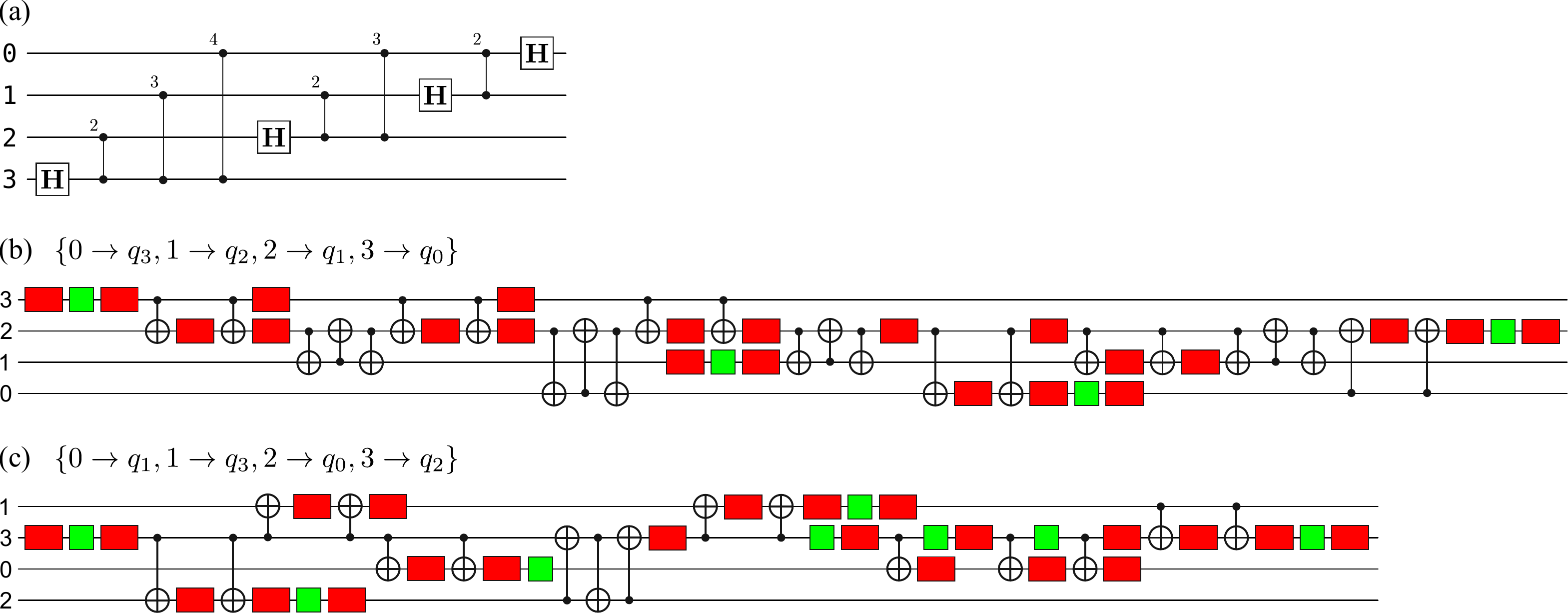}
\caption{Four-qubit QFT circuits (without the final swaps). (a) virtual circuit, (b) and (c) transpiled circuits with different layouts of the coupling maps $\{i\rightarrow q_j\}$ indicated above them; the red (green) boxes represent the $\mathbf{R}_z$ ($\mathbf{\sqrt{X}}$) gates.}
\label{fig:QFT_transpile}
\end{figure*}

\subsection{Native gates, connectivity, and circuit transpilation}
\label{subsec:transpile}

The native gates of a quantum processor are the set of gates that can be physically implemented. As shown in Table~\ref{tab:ibm_devices}, at ibm\_perth, this set consists of $\{\mathbf{I},\mathbf{X},\mathbf{\sqrt{X}},\mathbf{R}_z(\theta),\textsc{\textbf{cnot}}\}$. Although restricted, it is universal, so that any gate can be decomposed into a combination of its elements. For instance, the Hadamard gate is, up to a global phase, given by $\mathbf{R}_z(\pi/2)\mathbf{\sqrt{X}}\mathbf{R}_z(\pi/2)$; a controlled-phase gate, like those in the QFT circuit [Fig.~\href{fig:Pty_circuit}{\ref{fig:Pty_circuit}(c)}], between qubits $i$ and $j$ will be $\mathbf{CP}^\lambda_{ij}=\mathbf{R}_{zj}(\pi/2^\lambda)\textsc{\textbf{cnot}}_{ij}
\mathbf{R}_{zj}(-\pi/2^\lambda)\textsc{\textbf{cnot}}_{ij}\mathbf{R}_{zi}(\pi/2^\lambda)$. Thus, the restriction in the set of native gates implies that the physically implemented circuits will have, in general, a larger number of gates and larger depths, which is detrimental for the performance of the algorithms in noisy devices. 

Another important aspect of a quantum device is the connectivity between the physical qubits, which defines the qubit pairs on which two-qubit gates can be implemented. The coupling map of ibm\_perth has an \textsf{H}-like configuration. This means that there are only 6 pairs of connected qubits ($q_0\leftrightarrow q_1$, $q_1\leftrightarrow q_2$, $q_1\leftrightarrow q_3$, $q_3\leftrightarrow q_5$, $q_4\leftrightarrow q_5$, and $q_5\leftrightarrow q_6$), among the 21 possible connected pairs in an ideal fully-connected device. To apply two-qubit gates between non-connected qubits, one must implement one or more swap gates to move the qubit states of interest until they are in adjacent positions on the coupling map. As a consequence, both the number of \textsc{cnot}s and the depth of the circuit increase, which introduces more errors into the algorithm. The \textsc{cnot} gates are implemented with much higher error rates than the single-qubit gates in the quantum processor (see Table~\ref{tab:ibm_devices}). 

To handle limited connectivity and a set of native gates, one resorts to transpilation.\cite{WeaverBook} This is the process of rewriting the input circuit as a circuit compatible with these constraints. In a simplified description, the transpiler performs the following tasks: (i) maps the virtual qubits of the input circuit into the physical qubits of the coupling map; (ii) introduces the swap gates in the circuit to make it compatible with connectivity of the device; (iii) decomposes the input gates in terms of the native gates; (iv) optimizes the circuit by reducing its depth through the combination or elimination of quantum gates (especially the swaps). The optimization process is stochastic, so that an input circuit transpiled repeatedly may generate different outputs, especially for the more complex circuits. By transpiling the input several times and computing the depth and number of single- and two-qubit gates of the outputs, one can choose the circuit that presents the lower values.

Let us use a four-qubit QFT circuit (without the final swaps) to illustrate these issues in a case of interest for this work. Figure~\href{fig:QFT_transpile}{\ref{fig:QFT_transpile}(a)} shows the ``virtual'' circuit. Figures~\href{fig:QFT_transpile}{\ref{fig:QFT_transpile}(b)} and \href{fig:QFT_transpile}{\ref{fig:QFT_transpile}(c)} show the corresponding transpiled circuits for a nonoptimal and an optimal layout of the coupling map $\{i\rightarrow q_j\}$, respectively, where the red (green) boxes represent the $\mathbf{R}_z$ ($\mathbf{\sqrt{X}}$) gates. Accounting for the number of $\textbf{\textsc{cnot}}$, $\mathbf{R}_z$ and $\mathbf{\sqrt{X}}$ gates, as well as the depth, the nonoptimal circuit has $(24,22,4,42)$, while the optimal one has $(16,22,8,34)$. Table~\ref{tab:QFT_transpile} shows the accounting of these parameters for the optimal transpiled QFT circuits as a function of the number of qubits. We will return to this point later when discussing the results.

\begin{table}[b]
\centering
\caption{Number of gates and depth of the optimal transpiled QFT circuits at IBM devices as a function of the number of qubits.}
\label{tab:QFT_transpile}
\begin{ruledtabular}
\begin{tabular}{lccccc}
 & Three & Four & Five & Six & Seven \\
 & qubits & qubits &  qubits & qubits & qubits \\[1mm] 
\hline \\[-2.5mm]
\multicolumn{1}{c}{$\textbf{\textsc{cnot}}$} & 7 & 16 & 28 & 42 & 62 \\ 
\multicolumn{1}{c}{$\mathbf{R}_z$} & 15 & 22 & 38 & 64 & 84 \\ 
\multicolumn{1}{c}{$\mathbf{\sqrt{X}}$} & 7 & 8 & 13 & 26 & 31 \\
\multicolumn{1}{c}{Depth} & 20 & 34 & 50 & 68 & 84 \\
\end{tabular}
\end{ruledtabular}
\end{table}

\subsubsection{Transpilation of the circuits for ptychography}

The circuits for quantum ptychography shown in Fig.~\href{fig:Pty_circuit}{\ref{fig:Pty_circuit}(a)} can be divided into three subcircuits: (i) state preparation, (ii) intermediate measurements, and (iii) QFT followed by the measurement in the computational basis. To implement the method, we must ensure that for each of the $3n$ circuits, the subcircuits (i) and (iii) have the same configuration; otherwise, the requirements for testing a state estimation method would not be fulfilled. If one transpiles each circuit individually, most likely, this will not occur given the stochastic nature of the process. To avoid this, we first transpile subcircuits (i) and (iii) individually several times and select the optimal one (e.g., Table~\ref{tab:QFT_transpile} shows the parameters of an optimal QFT); the subcircuit (ii) has only single-qubit gates acting on one qubit at a time [Fig.~\href{fig:Pty_circuit}{\ref{fig:Pty_circuit}(b)}], so that a single transpilation is sufficient. After that, we build the $3n$ circuits by concatenating the three optimal subcircuits.

\subsection{Measurement error mitigation}
\label{subsec:ErrorMitigation}

The current quantum computers are severely affected by noise from different sources such as gate infidelity, measurement errors (see Table~\ref{tab:ibm_devices}), and decoherence.\cite{Abhijith2022} There are different approaches to handling this noise, and a prominent one is called error mitigation, where sophisticated and robust techniques have been developed recently.\cite{Bravyi2021,Nation2021,Kim2023} Addressing some of those techniques is beyond the scope of this work; however, it might be instructive to implement a simple one in our study. With this purpose, we consider the following measurement error mitigation technique. First, $2^n$ $n$-qubit circuits are built, where in each of them a state of the computational basis $\{|j\rangle\}$ is prepared and also measured in the computational basis $\{|j'\rangle\}$. Each circuit is executed $N_s$ times and gives us a distribution, $\Omega_j$, of $N_s$ experimental outcomes (shots). By normalizing each shot distribution, we generate a vector of conditional probabilities $\vec{P}_j=\{p(j'|j)\}_{j'=0}^{2^n-1}$ that will form the columns of a $2^n\times 2^n$ calibration matrix, $M$; note that in the absence of noise, we would have $p(j'|j)=\delta_{jj'}$ and $M$ would be an identity matrix. Now, let $\Omega$ be the shot distribution resulting from $N_s$ runnings of an $n$-qubit circuit where a given algorithm was carried out.  Using the calibration matrix, we can generate a new shot distribution by $\Omega'= M^{-1}\Omega$, which hopefully will have the measurement errors mitigated.

\subsubsection{Readout error mitigation on the ptychographic data}

Due to the intermediate measurement on a single qubit, each circuit for quantum ptychography, shown in Fig.~\href{fig:Pty_circuit}{\ref{fig:Pty_circuit}(a)}, generates a $2^{n+1}$-entry vector of shots, 
$\Omega_{\xi q}= \{|\langle j|\mathbf{U}|\psi_{\xi q}^\pm\rangle|^2\}_{j=0}^{2^n-1}$, 
where $\xi=x,y,z$ and $q=0,\ldots, n-1$. Thus, we have a $2^{n+1}\times 2^{n+1}$ global calibration matrix built as the tensor product of the single-qubit calibration matrices with the $n$-qubit one. We then apply the method described above to the outcomes generated in each circuit and obtain the set of ptychographic data with mitigated readout errors, $\Omega_{\xi q}'$.

\section{Results}
\label{sec:Results}

\subsection{PIE algorithm}  \label{subsec:PIE}

The PIE algorithm is the postprocessing mechanism that will estimate the state from the ptychographic data $\{\sqrt{\Omega_{\ell}}\}$, the used set of projectors $\{\bm{\Pi}_{\ell}\}$ and the basis for the final projective measurement, defined by the unitary $\mathbf{U}$. As shown in Algorithm~\ref{alg:PIE}, it starts with a random estimate of the input state that is updated throughout the loops and provides a final estimate when some stopping criterion is met. Here, we give further details on how the algorithm was operated in this work and also show an approach that improved its convergence compared to previous approaches.\cite{Fernandes2019,Fernandes2020} 

\subsubsection{Analysis of results}

In Algorithm~\ref{alg:PIE}, a PIE iteration consists of completing the loops over the $N$ projectors. As a stopping condition, we used here a fixed number of PIE iterations. To analyze the convergence of the algorithm, at each PIE iteration, we compute the trace distance between the normalized versions of the current ($|\varphi_\textrm{curr}\rangle$) and updated ($|\varphi_\textrm{updt}\rangle$) estimates of the states, defined as \cite{NielsenBook}
\begin{equation} \label{eq:TraceDistance}
D(|\varphi_\textrm{curr}\rangle,|\varphi_\textrm{updt}\rangle) = \sqrt{1 - |\langle\varphi_\textrm{curr}|\varphi_\textrm{updt}\rangle|^2},
\end{equation}
where $0\leqslant D\leqslant 1$. As the algorithm progresses, $D$ will decrease as the estimates become closer (or less distinguishable), and a perfect convergence would imply $D=0$. This measure does not assume any knowledge of the state being estimated. However, to evaluate the method, we must know it in advance so that it is possible to determine the quality of the estimations. To do this, at each PIE iteration, we also compute the fidelity between the input state and $|\varphi_\textrm{updt}\rangle$, defined as
\begin{equation} \label{eq:Fidelity}
F(|\psi\rangle,|\varphi_\textrm{updt}\rangle)=|\langle\psi|\varphi_\textrm{updt}\rangle|^2,
\end{equation}
where $0\leqslant F\leqslant 1$ and a perfect estimation leads to $F=1$. 

In addition to estimating the state, distances, and fidelities, another piece of data that we will obtain from the algorithm is the total execution time, so that it can be compared with the postprocessing algorithms of the Pauli and PZD  methods (see Sec.~\ref{subsec:PauliPZD}) under this criterion.

\subsubsection{A variable feedback parameter}

The feedback parameter $\beta$, shown in Algorithm~\ref{alg:PIE}, controls the step-size of the update. Usually, both classical\cite{Rodenburg04,Bunk2008,Maiden2009} and quantum\cite{Fernandes2019,Fernandes2020} ptychography use a constant value of $\beta$ found empirically in the range $(0,2]$. We began our study by simulating the ptychographic method (see discussion below), where the PIE algorithm was executed using a constant value of $\beta=1.5$, following the recipe of Ref.~\onlinecite{Fernandes2019}. Doing so, the results obtained were not as good as expected, even after trying other values for $\beta$. To circumvent this, we used the fact that the PIE algorithm is equivalent to an alternate gradient descent method.\cite{Fernandes2019} In this sense, $\beta$ can be identified as the learning rate, and a mechanism to drive the convergence of the algorithm to a global minimum is to make the learning rate to vary as the iterations proceed. Therefore, after several tests, we adopted a parameter $\beta$ that decreases with iterations. More specifically, the algorithm starts with $\beta=2$ and this value is decreased by $\Delta\beta$ at each PIE iteration. In this work we used either $\Delta\beta=0.1$ or $\Delta\beta=0.04$, which fixes the number of PIE iterations to $20$ or $50$, respectively. 

\begin{figure}[t]
\centerline{\includegraphics[width=1\columnwidth]{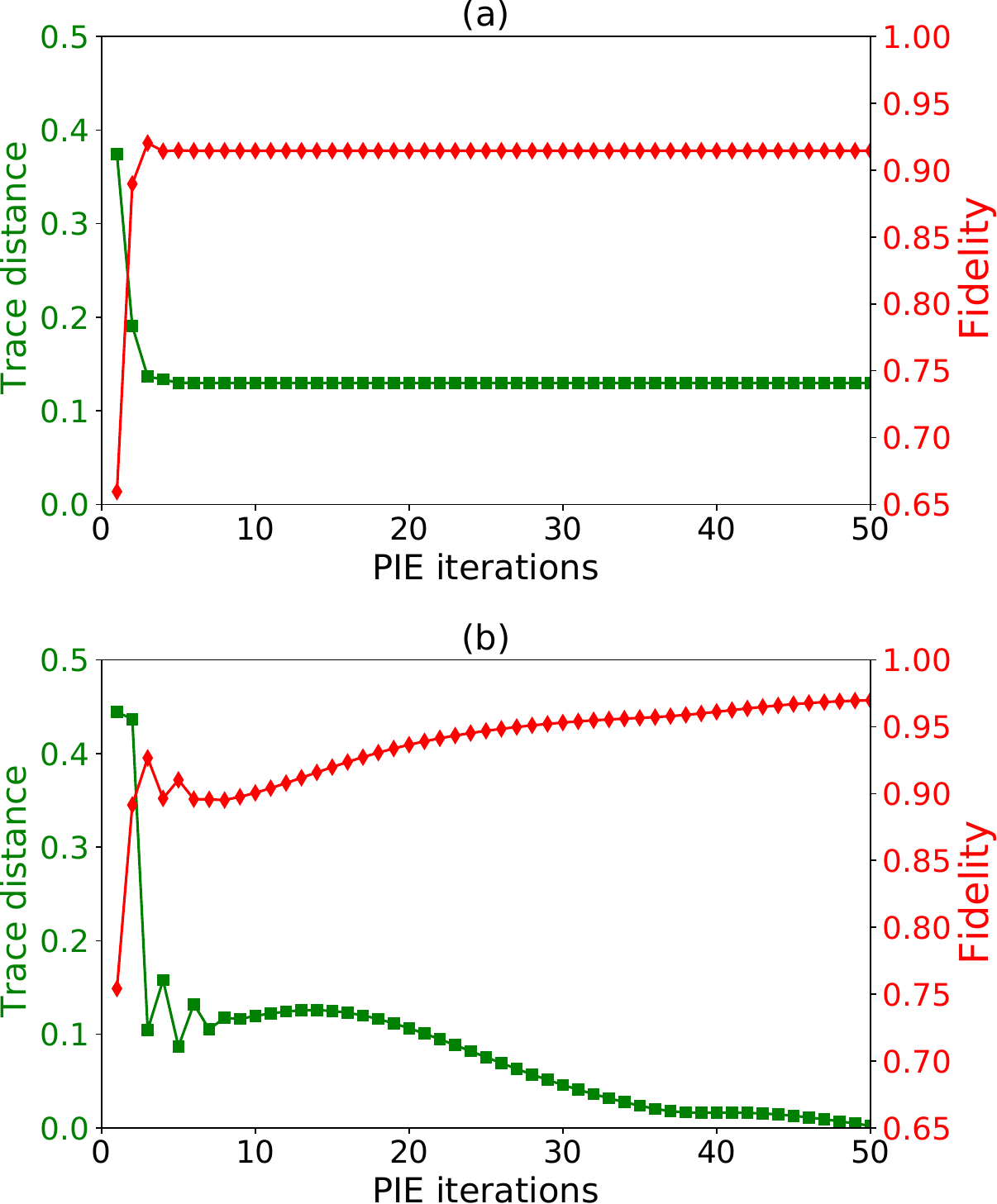}}
\caption{Progression of the PIE algorithm for estimating the Bell state $(|00\rangle+|11\rangle)/\sqrt{2}$ from experimental data. Trace distance and fidelity are plotted as a function of PIE iterations when the algorithm is run with the feedback parameter (a) fixed ($\beta=1.5$) and (b) variable ($\Delta\beta=0.04$).}
\label{fig:BetaParameter}
\end{figure}

\begin{figure*}[t]
\centerline{\includegraphics[width=1\textwidth]{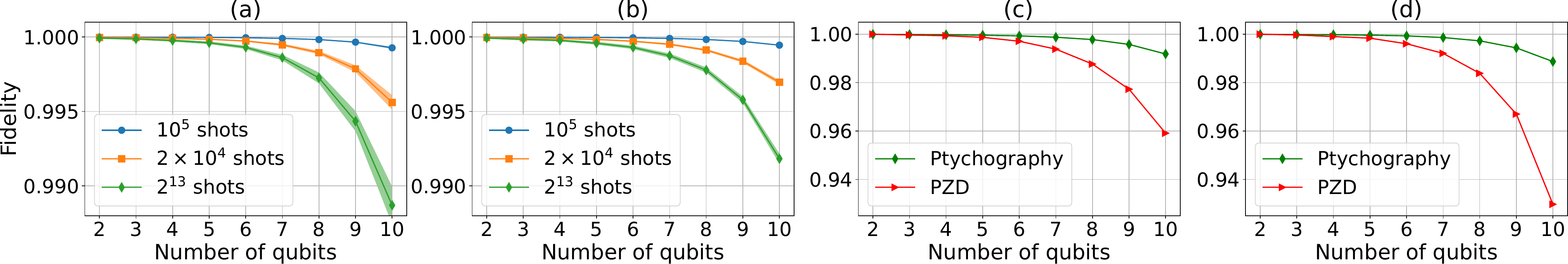}}
\caption{(a) and (b) Average fidelities for the ptychographic estimation of randomly generated separable and arbitrary states, respectively, as a function of the number of qubits. The simulations were performed with a fixed number of shots indicated at the insets; shaded areas represent the non-negligible standard deviations. (c) and (d) Comparison between the ptychographic estimations (green diamonds) using $2^{13}$ shots shown in (a) and (b), respectively, with the PZD estimations (red triangles) from simulations in equivalent conditions.\cite{Pereira2022}}
\label{fig:NoiselessFidelities}
\end{figure*}

To illustrate this discussion, we consider the estimation of the Bell state $|\psi\rangle=(|00\rangle+|11\rangle)/\sqrt{2}$. The corresponding  ptychographic data were generated experimentally (see more details in Sec.~\ref{subsec:ExperimentQFT}) and analyzed in the PIE algorithm without applying error mitigation. The plots in Figs.~\href{fig:BetaParameter}{\ref{fig:BetaParameter}(a)} and \href{fig:BetaParameter}{\ref{fig:BetaParameter}(b)} show the progression of the algorithm as a function of the PIE iterations for a fixed ($\beta=1.5$) and variable ($\Delta\beta=0.04$) feedback parameter, respectively. In Fig.~\href{fig:BetaParameter}{\ref{fig:BetaParameter}(a)}, one observes that the trace distance initially decreases but becomes stagnant after the fourth iteration, which also imposes a stagnation for the fidelity; in the end, these quantities reach the values $D=0.129$ and $F=0.915$, respectively. On the other hand, by varying $\beta$, it can be seen in Fig.~\href{fig:BetaParameter}{\ref{fig:BetaParameter}(b)} that there is an almost progressive decrease of the trace distance (despite some instability and small increase in the first iterations) with a simultaneous increase of the fidelity, showing the convergence of the algorithm; in the end they reached $D=0.002$ and $F=0.970$, respectively. Similar behaviors were observed in many other tests we performed for this work, from both simulated and experimental data. As this example attests, the difference between the approaches is significant, so the results presented hereafter were all obtained using a variable $\beta$ parameter in the PIE algorithm.

\subsection{Numerical simulations}
\label{subsec:NumericalSim}

Inspired by the work of PZD,\cite{Pereira2022} we performed simulations of the ptychographic method on a noise-free quantum computer, but considering a fixed number of shots, $N_s$. This number establishes a finite size for the ensemble used in the process, which introduces errors in the estimation of the count rates associated with the measurements and, consequently, in the estimation of the state. In addition to studying the method from this perspective, we can also compare our results with those of PZD.

The simulations were entirely carried out through Qiskit, using codes developed in Python at all stages, from creating the circuits to analyzing the results. To implement them, we adopted the following general procedures: (i) define the quantum states to be tested and for each one build the $3n$ circuits for ptychography as described in Sec.~\ref{subsec:transpile}; (ii) fix the number of shots, that is, the number of repetitions of each circuit; and (iii) feed the PIE algorithm with the generated ptychographic data and analyze the results. 

Similarly to PZD, for this study, we consider two sets of states: (i) random separable states, generated from random single-qubit gates, and (ii) random arbitrary states, generated from normalized vectors of $2^n$ random complex numbers. The difference between them is that the states in set (ii) will most likely be entangled states. The simulations were implemented for $n=2,3,\ldots,10$ qubits; for each $n$, we generated $100$ random states of each set and simulated the measurements with $N_s=2^{13}$ (number of shots in PZD's work), $N_s=2\times 10^4$ (an intermediate number) and $N_s=10^5$ (maximal number of shots allowed at ibm\_perth). The ptychographic data produced were analyzed in the algorithm with $20$ PIE iterations. For each state, we ran the algorithm $100$ times and computed the average fidelity; then, we calculated the average of these fidelities over the $100$ states and their standard deviation.

The results obtained for separable and arbitrary states are shown in Figs.~\href{fig:NoiselessFidelities}{\ref{fig:NoiselessFidelities}(a)} and \href{fig:NoiselessFidelities}{\ref{fig:NoiselessFidelities}(b)}, respectively, as a function of the number of qubits. In both, it is observed that, for a fixed number of shots, the fidelities decrease as the number of qubits increases, since the effect of the finite size of the ensemble is relative to the dimension of the quantum system. This effect can be mitigated by increasing the size of the ensemble, which can be seen by looking at these plots as a function of $N_s$. Two other conclusions can be drawn from these results: first, the standard deviations were, in general, negligible, indicating that the method generated very similar fidelities for all the states tested; second, a comparison between the plots in Figs.~\href{fig:NoiselessFidelities}{\ref{fig:NoiselessFidelities}(a)} and \href{fig:NoiselessFidelities}{\ref{fig:NoiselessFidelities}(b)} shows that the performance of quantum ptychography in a noise-free device does not differ significantly for separable and entangled states. 

Our results using $2^{13}$ shots can be compared with those from PZD. In particular, we will consider their best case where $4n+1$ separable bases were used to implement the method. In Figs.~\href{fig:NoiselessFidelities}{\ref{fig:NoiselessFidelities}(c)} and \href{fig:NoiselessFidelities}{\ref{fig:NoiselessFidelities}(d)}, we plot the average fidelities of the ptychographic method (green diamonds) and the median fidelities of the PZD method (red triangles; extracted from Ref.~\onlinecite{Pereira2022}) for separable and arbitrary states, respectively. The performances of both methods are similar up to $n=5$ qubits, but quantum ptychography outperformed PZD for larger values of $n$. For example, for $n=10$, we obtained average fidelities of $0.989$ for separable states and $0.992$ for arbitrary states, while the median fidelities for these sets obtained by PZD were $0.959$ and $0.930$, respectively. Furthermore, unlike ptychography, the PZD method showed, more clearly, different performances considering the estimation of separable and entangled states, achieving better fidelities for the former, as can be seen by comparing the plots of Figs.~\href{fig:NoiselessFidelities}{\ref{fig:NoiselessFidelities}(c)} and \href{fig:NoiselessFidelities}{\ref{fig:NoiselessFidelities}(d)}. 

\begin{table*}[t]
\caption{Multiqubit target states used in the experiments.}
\label{tab:Qubit_states} 
\begin{ruledtabular}
\begin{tabular}{lllcl}
\multicolumn{5}{c}{Two-qubit states} \\[0.5mm]
\hline\\[-1.5mm]  
$\displaystyle|\psi_1\rangle=\left[\frac{|0\rangle+|1\rangle}{\sqrt{2}}\right]^{\otimes 2}$ & 
$\displaystyle|\psi_2\rangle=\left[\frac{|0\rangle-|1\rangle}{\sqrt{2}}\right]^{\otimes 2}$ & $\displaystyle|\psi_3\rangle=\left[\frac{|0\rangle+e^{i\pi/4}|1\rangle}{\sqrt{2}}\right]^{\otimes 2}$ & $\displaystyle|\psi_4\rangle=\left[\frac{|0\rangle-e^{i\pi/4}|1\rangle}{\sqrt{2}}\right]^{\otimes 2}$ & 
$\displaystyle|\psi_5\rangle=\frac{|00\rangle+|11\rangle}{\sqrt{2}}$  \\[5mm]
$\displaystyle|\psi_6\rangle=\frac{|00\rangle-|11\rangle}{\sqrt{2}}$ & $\displaystyle|\psi_7\rangle=\frac{|01\rangle+|10\rangle}{\sqrt{2}}$ & $\displaystyle|\psi_8\rangle=\frac{|01\rangle-|10\rangle}{\sqrt{2}}$ & \multicolumn{2}{l}{$|\psi_9\rangle$ and $|\psi_{10}\rangle=|\textrm{Random state}\rangle$} \\[4mm]
\hline \\[-2.75mm]
\multicolumn{5}{c}{$n$-qubit states} \\[0.75mm]
\hline\\[-1mm] 
$\displaystyle|\psi_1^n\rangle=\left[\frac{|0\rangle+e^{i\pi/4}|1\rangle}{\sqrt{2}}\right]^{\otimes n}$ & $\displaystyle|\psi_2^n\rangle=\left[\frac{|0\rangle-e^{i\pi/4}|1\rangle}{\sqrt{2}}\right]^{\otimes n}$ & 
$\displaystyle|\psi_3^n\rangle=|\textrm{Random separable}\rangle$ 
& $|\psi_4^n\rangle=|\textrm{GHZ}\rangle_n$  & 
$|\psi_5^n\rangle=|W\rangle_n$ \\[3mm]
\end{tabular}
\end{ruledtabular}
\end{table*}

\begin{figure}[b]
\centerline{\includegraphics[width=1\columnwidth]{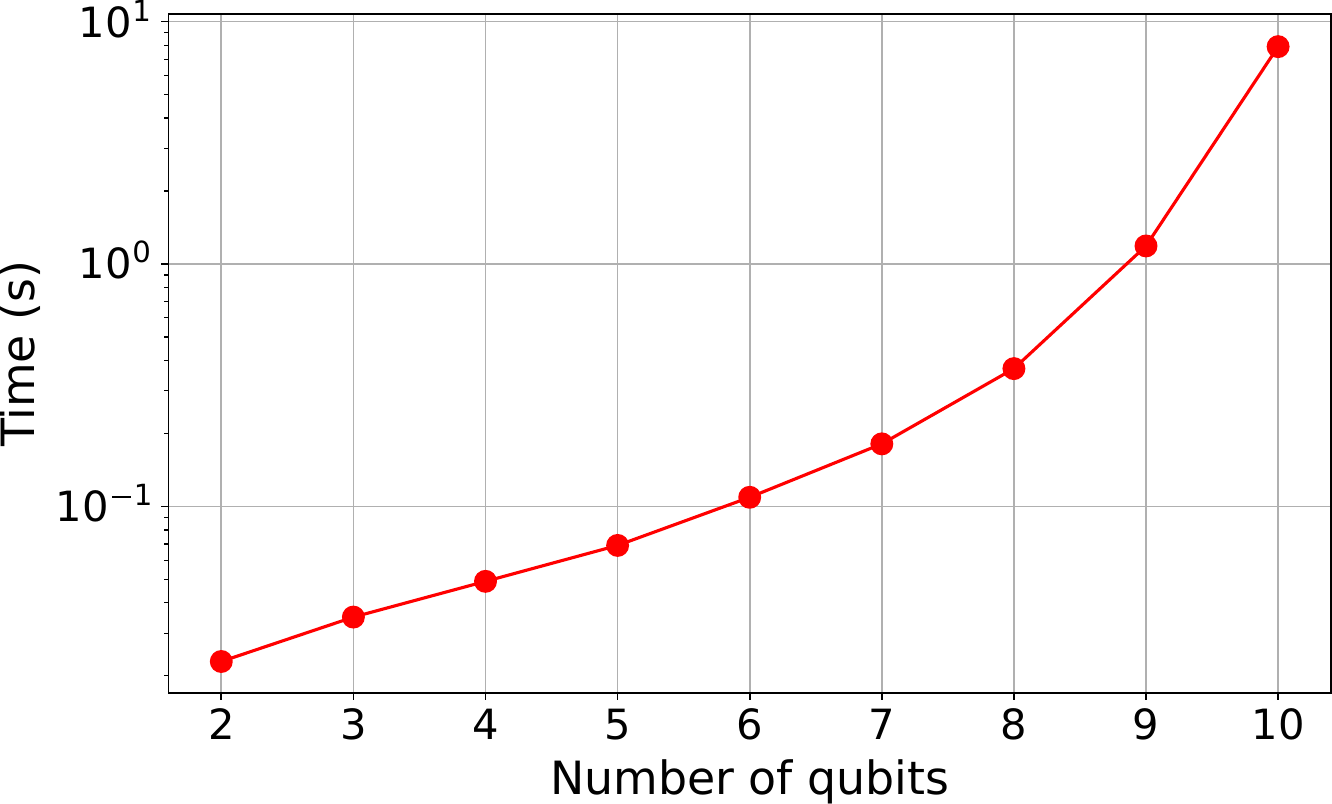}}
\caption{Average times as a function of the number of qubits that the PIE algorithm takes to process the input ptychographic data and output an estimate; these times were obtained for 20 PIE iterations. Standard deviations were negligible and are not noticeable.}
\label{fig:PtyTIME}
\end{figure}

In our simulations, performed on a portable personal computer, we also determined the average times that the PIE algorithm takes to process the input ptychographic data and output an estimate. The results obtained are shown in Fig.~\ref{fig:PtyTIME} as a function of $n$. For ten qubits, the required time was, on average, $8$~s, which is quite fast when compared to the fastest maximum likelihood estimation algorithm used in standard tomography that takes about $4.7\times 10^3\!$~s;\cite{Shang2017} on the other hand, the postprocessing times of the PZD method were shown to be even faster, achieving $\sim 0.8\!$~s. It is important to highlight that the processing times of the PIE algorithm may vary for a different number of iterations or a different stopping condition. In addition, there is a way to reduce these times by improving the algorithm itself: here we used the version of Ref.~\onlinecite{Fernandes2019}, which is adapted from the simplest form of the algorithm proposed in 2004;\cite{Rodenburg04} since then, it has undergone many improvements,\cite{Maiden2017} which could also be adapted to quantum ptychography and remain a topic for future investigations.

\subsection{Experimental results}
\label{subsec:ExperimentQFT}

The ptychographic method was implemented on a quantum processor through the following steps:
\begin{enumerate}
\item[(1)] Define the quantum state to be tested and build the $3n$ circuits for ptychography as described in Sec.~\ref{subsec:transpile}. 
\item[(2)] Carry out the experiment through online access to the device ibm\_perth, using the maximum allowed number of shots per circuit, $N_s=10^5$ . 
\item[(3)] Duplicate the generated ptychographic data and subject one set to the measurement error mitigation (Sec.~\ref{subsec:ErrorMitigation}). 
\item[(4)] Feed both sets into the PIE algorithm, which is ran with $50$ PIE iterations (i.e., $\Delta\beta=0.04$), and compute the average fidelities and standard deviations across $100$ runs.
\end{enumerate}
We used a fixed but diverse set of multiqubit states, containing both separable and entangled ones, as shown in Table~\ref{tab:Qubit_states}. The results obtained will be presented separately for $n=2$ and $n>2$ qubits.

\subsubsection{Results for \texorpdfstring{$n=2$}{n=2} qubits} 

\begin{figure}[t]
\centerline{\includegraphics[width=1\columnwidth]{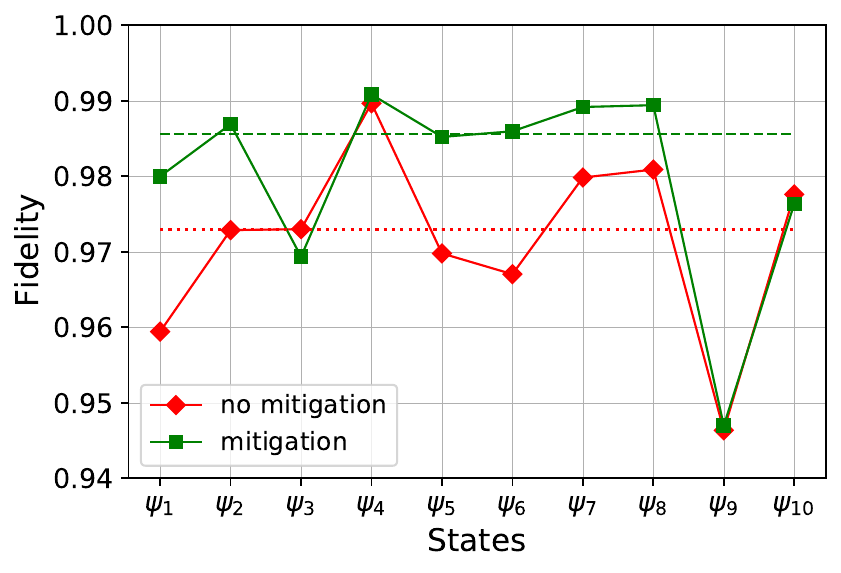}}
\caption{Fidelities for the experimental ptychographic estimation of the two-qubit states $\{|\psi_j\rangle\}_{j=1}^{10}$ shown in Table~\ref{tab:Qubit_states}. Red diamonds and green squares correspond to the results achieved without and with measurement error mitigation, respectively. The red dotted and green dashed lines correspond to the median fidelity in each case. Standard deviations were negligible and are not noticeable in the plots.}
\label{fig:2qbExperiments}
\end{figure}

In Table~\ref{tab:Qubit_states}, we list the set of ten two-qubit states used in the experiments. In this set, we have separable ($|\psi_1\rangle$ to $|\psi_4\rangle$) and entangled states, including the four Bell states ($|\psi_5\rangle$ to $|\psi_8\rangle$) and arbitrary random states  ($|\psi_{9}\rangle$ and $|\psi_{10}\rangle$). Figure~\ref{fig:2qbExperiments} shows the fidelities, $F$ and $F'$, achieved in their estimation without (red diamonds) and with (green squares) the application of measurement error mitigation, respectively; red dotted and green dashed lines correspond to the median fidelity ($\tilde{F}$ and $\tilde{F}'$) in each case. In both cases, the fidelities ranged from $\sim\!0.947$ to $\sim\!0.99$, but there was a clear improvement for most states when we applied the error mitigation. The medians before and after this were $\tilde{F}=0.973$ and $\tilde{F}'=0.986$, respectively. Overall, the achieved fidelities were high, and there was a balance in the quality of the estimations considering the separable and entangled states; the four Bell states, for instance, reached fidelities $F'_\textrm{Bell}>0.985$, while the separable (entangled) state $|\psi_4\rangle$ ($|\psi_9\rangle$) reached the highest (lowest) fidelity $F_4'=0.991$ ($F_9'=0.947$).

\subsubsection{Results for \texorpdfstring{$n>2$}{n>2} qubits} 

\begin{figure*}[htb]
\centerline{\includegraphics[width=1\textwidth]{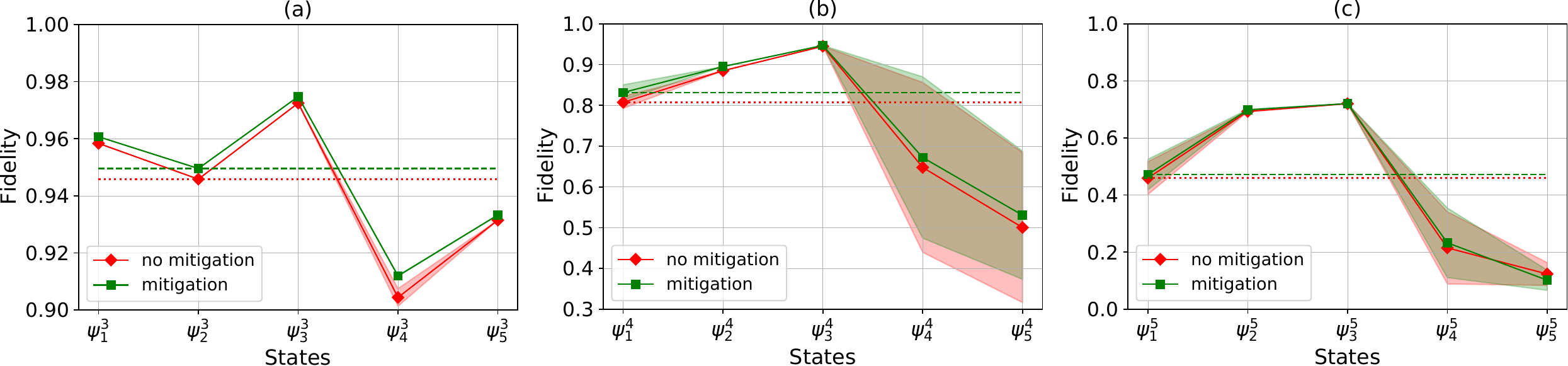}}
\caption{Fidelities for the experimental ptychographic estimation of the $n$-qubit states $\{|\psi_j^n\rangle\}_{j=1}^{5}$ shown in Table~\ref{tab:Qubit_states}, using the QFT: (a) $n=3$, (b) $n=4$, and (c) $n=5$. Red diamonds and green squares correspond to the results achieved without and with measurement error mitigation, respectively; shaded areas represent the non-negligible standard deviations. The red dotted and green dashed lines correspond to the median fidelity in each case. }
\label{fig:nqbExperiments}
\end{figure*}

In Table~\ref{tab:Qubit_states}, we list the set of five $n$-qubit states used in the experiments. In this set, we selected separable states ($|\psi_1\rangle$ to $|\psi_3\rangle$), and the entangled states $|\psi_4\rangle=|\textrm{GHZ}\rangle=(|0\rangle^{\otimes n}+|1\rangle^{\otimes n})/\sqrt{2}$ and $|\psi_5\rangle=|W\rangle=(|100\ldots 0\rangle+|010\ldots 0\rangle+\cdots+|000\ldots 1\rangle)/\sqrt{n}$. They have been estimated for $n=3,4$, and $5$ qubits. The fidelities are shown in Figs.~\href{fig:nqbExperiments}{\ref{fig:nqbExperiments}(a)}, \href{fig:nqbExperiments}{\ref{fig:nqbExperiments}(b)}, and \href{fig:nqbExperiments}{\ref{fig:nqbExperiments}(c)}, respectively, and were obtained without (red diamonds) and with (green squares) the application of measurement error mitigation; the shaded areas represent the standard deviation. In these plots, the red dotted and green dashed lines correspond to the median fidelity ($\tilde{F}$ and $\tilde{F}'$) in each case. 

The effect of error mitigation leading to an increase in the fidelity is still noticeable for most states but it becomes less and less relevant as the number of qubits increases.  For $n=3$ qubits [Fig.~\href{fig:nqbExperiments}{\ref{fig:nqbExperiments}(a)}], we obtained good fidelities in all cases; the medians without and with error mitigation were $\tilde{F}=0.946$ and $\tilde{F}'=0.95$, respectively. There is a clear superiority in the estimation of separable states over the entangled ones: the former reached fidelities $F'_\textrm{sep}>0.949$, whereas for the latter, $F'_\textsc{ghz}=0.912$ and $F_W'=0.933$, which were high taking into account that they are much more affected by preparation errors. On the other hand, for $n=4$ qubits [Fig.~\href{fig:nqbExperiments}{\ref{fig:nqbExperiments}(b)}], only the separable states were estimated with acceptable fidelities ($F'_\textrm{sep}>0.83$), while the estimations of the entangled states presented poor results, thanks to the errors introduced by the combination of the preparation and QFT circuits. Finally, for $n=5$ qubits [Fig.~\href{fig:nqbExperiments}{\ref{fig:nqbExperiments}(c)}], not even the separable states showed good results, despite being estimated with fidelities far greater than those of entangled states. In this case, the noise of the QFT circuit alone was enough to overthrow the fidelities of the separable states.

Given the scenario reported above, in the next section, we will look for different approaches to the ptychographic method that could be more suitable for noisy quantum devices. The experimental results presented here will be further discussed and compared to those from PZD in Sec.~\ref{subsec:OverviewExp}.

\section{Multiqubit ptychography with other unitary operations}
\label{sec:ExperimentAQFT}

As described in Sec.~\ref{subsec:Ptycho_overview}, the ptychographic method works for a suitable choice of the intermediate overlapping projections $\{\bm{\Pi}_\ell\}$ and the basis for the final measurement, defined by a unitary operation $\mathbf{U}$. So far, we carried out the original proposal,\cite{Fernandes2019} with the projections given by Eq.~(\ref{eq:Pi_xiq}) and $\mathbf{U}$ given by the QFT. The experimental results for $n>3$ qubits indicated that, with this choice, the method will face scalability issues for the current noisy quantum devices. In an attempt to circumvent this, we investigated the use of different unitary operations combined with the \emph{same} intermediate projections of Eq.~(\ref{eq:Pi_xiq}). This study was not intended to be exhaustive but only to point out possible paths that could lead to the implementation of the ptychographic method in a scalable way on noisy devices.

\subsection{Approximate quantum Fourier transform}   \label{subsec:AQFT}

A natural alternative to the QFT, with an operation close to it but with simpler circuits, is the approximate quantum Fourier transform (AQFT).\cite{Coppersmith1994} Next, we briefly describe this operation and present the results of its use in the ptychographic method. 

\subsubsection{An overview of the AQFT}

The AQFT for $n$ qubits is characterized by a parameter $m=1,\ldots,n$, which sets the degree of approximation. In the regular $n$-qubit QFT circuit, we have the controlled-phase gates with indices $k=2,\ldots,n$, as shown in Fig.~\href{fig:Pty_circuit}{\ref{fig:Pty_circuit}(c)}. The effect of the AQFT of degree $m$ in this circuit is to remove such gates with $k>m$.\cite{Coppersmith1994} For $m=1$, all controlled gates are eliminated, and only one Hadamard gate remains on each qubit (Hadamard transform, $\mathbf{H}^{\otimes n}$). For $1<m<n$, we have approximations of the QFT with controlled gates only between the qubits separated by distances smaller than $m$ in the circuit. For $m=n$, we have the exact QFT. 

It is important to highlight that the smaller $m$, the lower the accuracy of the approximation. However, in the presence of decoherence, a lower accuracy may result in a better performance of the operation.\cite{Barenco1996} The approximation reduces the number of controlled-phase gates to $(2n-m)(m-1)/2$, by removing those ones that act on qubits that are most far apart in the circuit. Regarding noise effects, this is beneficial for devices with limited connectivity and set of native gates: there is a reduction in the circuit depth and in the number of \textsc{cnot}s (including those forming the swap gates that connect the ``disconnected'' qubits), thus reducing noise. Considering the AQFT of degree $m=2$, which will be implemented in our experiments, Table~\ref{tab:AQFT_transpile} shows the accounting of these parameters for its optimal transpiled circuit as a function of the number of qubits. A comparison with the numbers of the QFT circuits shown in Table~\ref{tab:QFT_transpile} clearly shows the reduction mentioned above.

\begin{table}[b]
\centering
\caption{Number of gates and depth of the optimal transpiled AQFT circuits at IBM devices as a function of the number of qubits for an AQFT of degree $m=2$.}
\label{tab:AQFT_transpile}
\begin{ruledtabular}
\begin{tabular}{lccccc}
& Three & Four & Five & Six & Seven \\
 & qubits & qubits &  qubits & qubits & qubits \\[1mm]  
\hline \\[-2.5mm]
\multicolumn{1}{c}{$\textbf{\textsc{cnot}}$} & $4$ & $6$ & $8$ & $11$ & $14$  \\ 
\multicolumn{1}{c}{$\mathbf{R}_z$} & $8$ & $11$ & $14$ & $22$ & $30$  \\ 
\multicolumn{1}{c}{$\mathbf{\sqrt{X}}$} & $3$ & $4$ & $5$ & $10$ & $15$  \\
\multicolumn{1}{c}{Depth} & $15$ & $21$ & $27$ & $37$ & $48$  \\
\end{tabular}
\end{ruledtabular}
\end{table}

\subsubsection{Numerical simulations}

To evaluate the feasibility of the AQFT for quantum ptychography, we first performed simulations in a noise-free quantum computer using the $n$-qubit states shown in Table~\ref{tab:Qubit_states} and approximations of degree $m=1,2,3,$ and $4$. The data were generated with $2\times 10^4$ shots and the PIE algorithm was run with $\Delta\beta=0.04$; for each state we computed the average fidelity and its standard deviation across ten runs. The unitary operation for the AQFT of degree $m$ to be used in the PIE algorithm (Algorithm~\ref{alg:PIE}) is given by \begin{align}    \label{eq:AQFT}
\mathbf{U}\rightarrow\mathbf{U}_\textsc{aqft}^m=\frac{1}{\sqrt{2^n}}\sum_{j,k=0}^{2^n-1}e^{2\pi i Y_{jk}^m/2^n}|j\rangle\langle k|,
\end{align}
where $Y_{jk}^m=\sum_{a,b=0}^{n-1}j_ak_b2^{a+b}$, with the sum restricted to $n-m\leqslant a+b\leqslant n-1$.

\begin{figure}[b]
\centerline{\includegraphics[width=1\columnwidth]{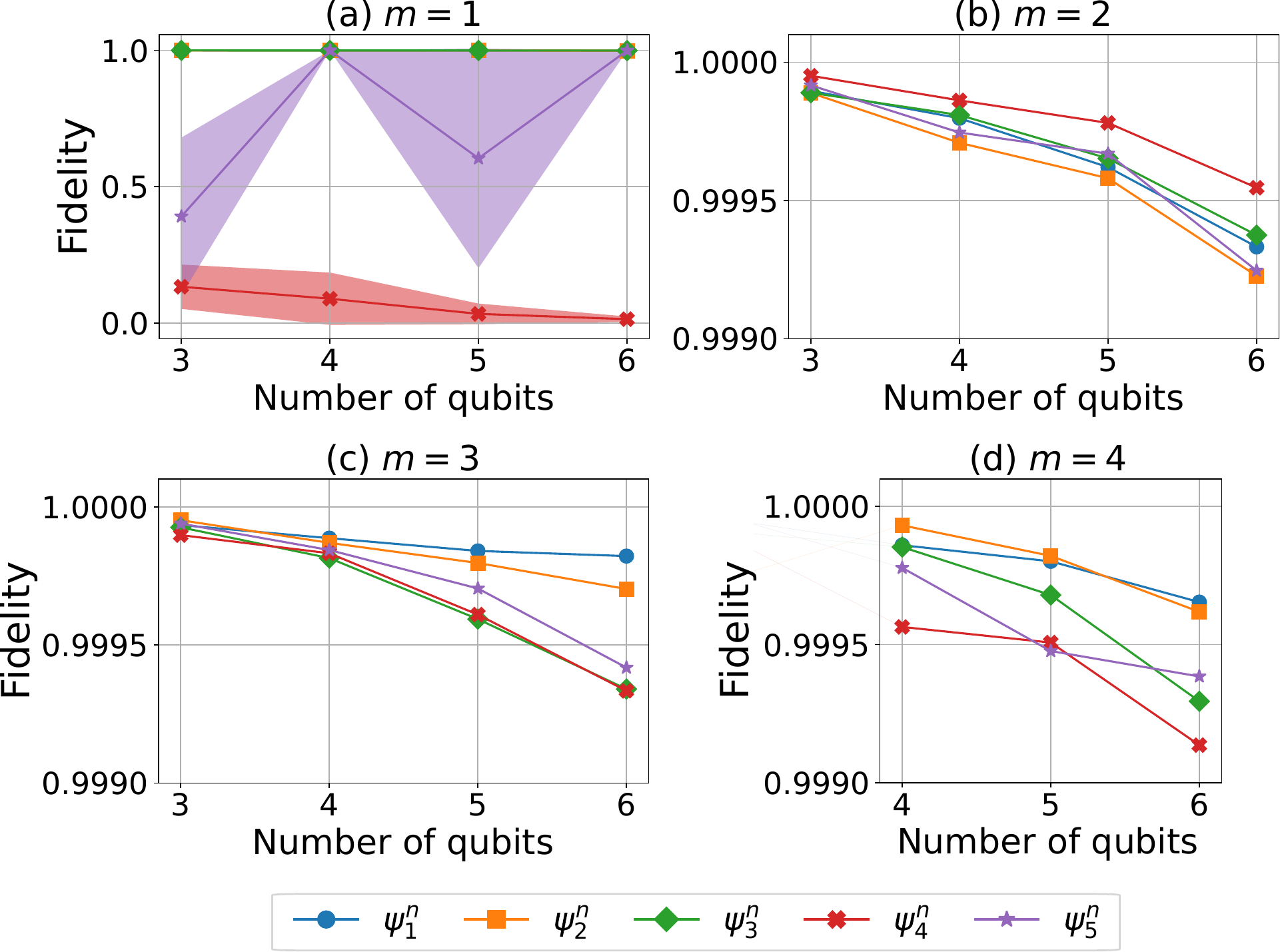}}
\caption{Fidelities of the $n$-qubit states $\{|\psi_j^n\rangle\}_{j=1}^{5}$ shown in Table~\ref{tab:Qubit_states}, estimated by the ptychographic method from simulations using the AQFT of degree $m$ shown above each plot. Shaded areas represent the non-negligible standard deviations. }
\label{fig:AQFTsimulations}
\end{figure}

Figure~\ref{fig:AQFTsimulations} shows the results obtained. For $m=1$, the estimations were nearly perfect for the separable states and very poor for the entangled states, especially the GHZ ($|\psi^n_4\rangle$); the $W$ state ($|\psi^n_5\rangle$) was correctly estimated only for an odd number of qubits, but we cannot explain this behavior. Therefore, the Hadamard transform is not a good candidate to replace the QFT, considering the intermediate projectors $\{\bm{\Pi}_{\xi q}^\pm\}$ [Eq.~(\ref{eq:Pi_xiq})]. On the other hand, AQFTs of degree $m>1$ showed nearly perfect estimations in all cases. The nonunit fidelities and their decrease with increasing $n$ can be attributed to the finite size of the ensemble. Therefore, AQFT with $m>1$ has the potential to replace the QFT in this scenario of quantum ptychography, but a more robust demonstration is needed.

\subsubsection{Experimental results}

\begin{figure*}[htb]
\centerline{\includegraphics[width=1\textwidth]{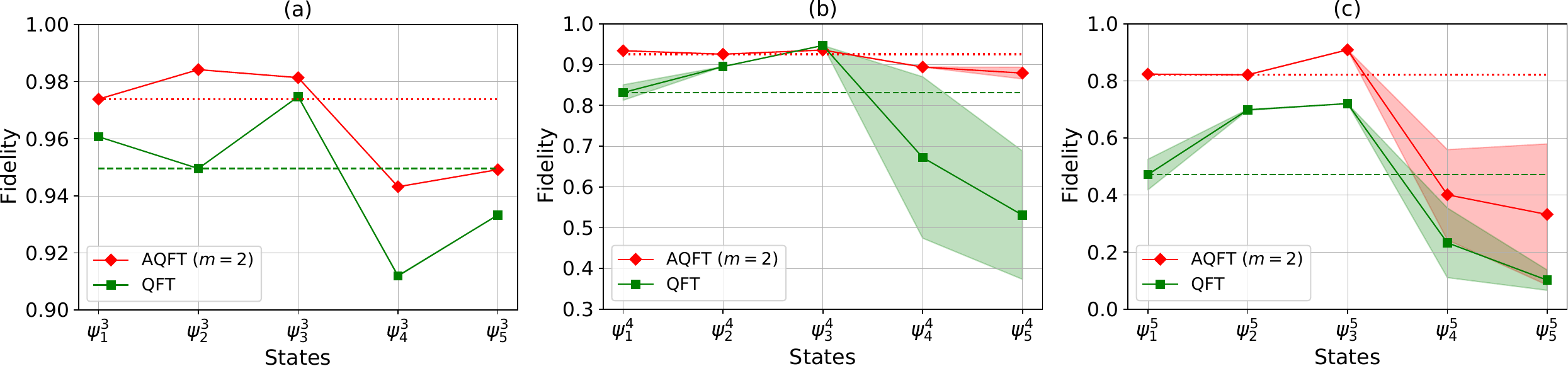}}
\caption{Fidelities for the experimental ptychographic estimation of the $n$-qubit states $\{|\psi_j^n\rangle\}_{j=1}^{5}$ shown in Table~\ref{tab:Qubit_states}, using the QFT (green squares) and AQFT of degree $m=2$ (red diamonds): (a) $n=3$, (b) $n=4$, and (c) $n=5$. Both cases correspond to the results achieved with measurement error mitigation; shaded areas represent the non-negligible standard deviations. The red dotted and green dashed lines correspond to the median fidelity in each case. }
\label{fig:AQFTexperiment}
\end{figure*}

Following the steps (1)--(4) described in Sec.~\ref{subsec:ExperimentQFT}, we performed the experiments using the AQFT of degree $m=2$. Here, we present only the results obtained with measurement error mitigation; once again, it had little effectiveness. For $n=3,4,$ and $5$, the $n$-qubit states shown in Table~\ref{tab:Qubit_states} were estimated with the fidelities shown by the red diamonds in Figs.~\href{fig:AQFTexperiment}{\ref{fig:AQFTexperiment}(a)}--\href{fig:AQFTexperiment}{\ref{fig:AQFTexperiment}(c)}, respectively. For the sake of comparison, we also plot the fidelities from Fig.~\ref{fig:nqbExperiments} obtained with the QFT and applying error mitigation (green squares). The shaded areas represent the standard deviation, and the red dotted and green dashed lines correspond to the median fidelities $\tilde{F}'_\textsc{aqft}$ and $\tilde{F}'_\textsc{qft}$, respectively.

It is observed that, with the exception of $|\psi_3^4\rangle$ (which had already shown high fidelity with the QFT), all other states were estimated with higher fidelities using the AQFT. This can also be seen when comparing the medians. The discrepancies in the estimation between separable $(|\psi_{1,2,3}^n\rangle)$ and entangled states $(|\psi_{4,5}^n\rangle)$ remain, as a consequence of much higher preparation errors for the latter. For $n=3$ qubits [Fig.~\href{fig:AQFTexperiment}{\ref{fig:AQFTexperiment}(a)}], the good fidelities achieved with the QFT were further improved with the AQFT ($\tilde{F}'_\textsc{qft}=0.95$ and $\tilde{F}'_\textsc{aqft}=0.974$). We highlight the results for $n=4$ qubits [Fig.~\href{fig:AQFTexperiment}{\ref{fig:AQFTexperiment}(b)}], where now all states presented good estimation fidelities with the AQFT. For the separable states, we have $F'_\textrm{sep}>0.925$ while for the entangled ones, $F'_\textsc{ghz}=0.894$ and $F'_W=0.879$, which represent a significant improvement over the QFT. Finally, for $n=5$ qubits [Fig.~\href{fig:AQFTexperiment}{\ref{fig:AQFTexperiment}(c)}], despite the improvement in the estimation of the entangled states, only the separable ones presented reasonably good fidelities ($F'_\textrm{sep}>0.82$). This shows the effectiveness of using AQFT, which strongly attenuated the noise of the ptychographic circuit, allowing good estimates of the separable states. The poor results for the entangled states, in this case, can be almost entirely accounted for by the noise in their preparation circuits.

In summary, the use of AQFT with $m>1$ showed promising results in the  simulations and, for $m=2$, better experimental estimations of up to five qubits in comparison with the QFT. However, for large $n$, even this lower-degree approximation may lead to circuits that become prohibitively noisy in the current quantum devices. Next, we will discuss the less noisy alternative, which consists of using separable unitary operations.

\subsection{Random separable unitary operations}

The AQFT of degree $m=1$, namely, the Hadamard transform, involves a separable operation $\mathbf{H}^{\otimes n}$, but the simulations showed that it is not suitable for estimating entangled states in the ptychographic protocol addressed here [see Fig.~\href{fig:AQFTsimulations}{\ref{fig:AQFTsimulations}(a)}]. As the measurements in separable bases require much simpler circuits (lower depths and no use of two-qubit gates), it is worth investigating whether there are other separable unitaries suitable for this protocol. For this purpose, we considered random operations over $n$ qubits given by
\begin{equation}    \label{eq:Urandom_sep}
   \mathbf{U}\rightarrow \mathbf{U}_\textrm{sep} = \mathbf{U}^\textrm{rnd}_{0} \otimes \mathbf{U}^\textrm{rnd}_{1} \otimes\cdots\otimes \mathbf{U}^\textrm{rnd}_{n-1},
\end{equation}
where  $\mathbf{U}^\textrm{rnd}_l\equiv\mathbf{U}(\theta_l,\phi_l,\lambda_l)= e^{i\frac{\phi_l+\lambda_l}{2}}\mathbf{R}_z(\phi_l)\mathbf{R}_y(\theta_l)\mathbf{R}_z(\lambda_l)$ is an arbitrary single qubit unitary operation made random by randomly generating the angles $(\theta_l,\phi_l,\lambda_l)$. In terms of the native gates of the IBM device, this unitary is given by
\begin{equation}
\mathbf{U}(\theta,\phi,\lambda)=\mathbf{R}_z(\lambda)\mathbf{\sqrt{X}} \mathbf{R}_z(\theta+\pi)\mathbf{\sqrt{X}}\mathbf{R}_z(\phi+\pi).
\end{equation}

\begin{figure}[b]
\centerline{\includegraphics[width=1\columnwidth]{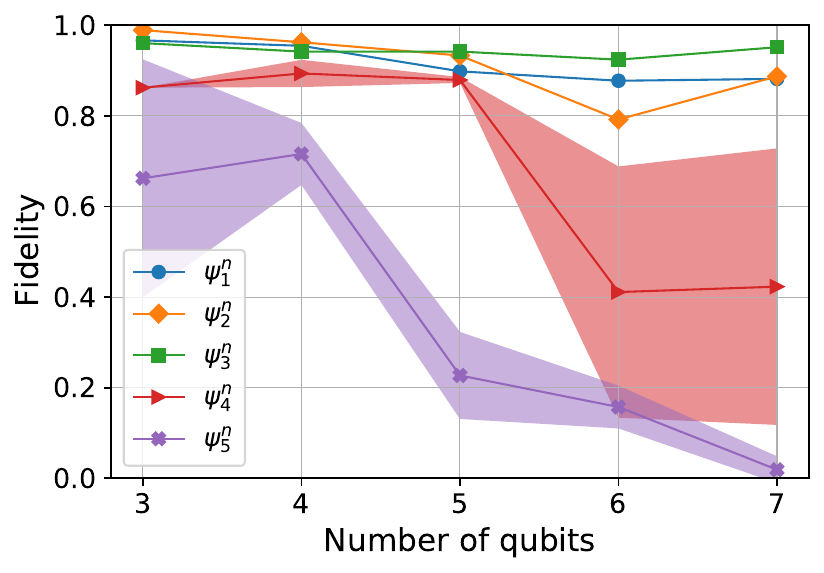}}
\caption{Fidelities for the experimental ptychographic estimation of the $n$-qubit states $\{|\psi_j\rangle\}_{j=1}^{5}$ shown in Table~\ref{tab:Qubit_states}, using random separable unitary operations. These results were achieved without applying measurement error mitigation; shaded areas represent the non-negligible standard deviations. }
\label{fig:Urandexperiment}
\end{figure}

Due to the low complexity of the circuits in the measurement stage, here we performed the experiment for up to $n=7$ qubits, the maximum at ibm\_perth. For each target state $\{|\psi_j^n\rangle\}_{j=1}^{5}$ shown in Table~\ref{tab:Qubit_states}, we generate random sets $\{(\theta_l,\phi_l,\lambda_l)\}_{l=0}^{n-1}$ defining the measurement basis through the unitary in Eq.~(\ref{eq:Urandom_sep}), which is also used in the PIE algorithm (Algorithm~\ref{alg:PIE}). The results obtained without applying measurement error mitigation are shown in Fig.~\ref{fig:Urandexperiment}. In general, the separable states presented high fidelities in all cases, for example, for $n=7$ qubits $F_\textrm{sep}>0.88$. On the other hand, the $W$ state $(|\psi_5^n\rangle)$ was poorly estimated in all cases, while the GHZ state $(|\psi_4^n\rangle)$ presented good fidelities only up to five qubits (its fidelity for $n=5$ was even much greater than in the previous experiments with QFT and AQFT, reaching $F_\textsc{ghz}=0.879$).

Although our test involved a restricted set of states, the results showed that using a separable measurement basis in combination with the also separable intermediate projections given by Eq.~(\ref{eq:Pi_xiq}) should work well only to estimate separable states. It does not seem to exist a separable basis in this scenario, which will make the ptychographic method work for arbitrary entangled states. In Sec.~\ref{subsec:FurtherSteps}, we discuss a possible approach to circumvent this issue.

\section{Discussion}
\label{sec:Discussion}

\subsection{An overview of the experimental results}
\label{subsec:OverviewExp}

Among the experimental results presented in Sec.~\ref{subsec:ExperimentQFT} and Sec.~\ref{subsec:AQFT}, we highlight the estimations of two- and three-qubit states using the QFT, as well as the estimations of three- and four-qubit states using the AQFT of degree $m=2$, where high fidelities were achieved in all cases. To put these results in perspective, we shall compare them with those obtained by PZD, which have also experimentally implemented their method on an IBM quantum processor (ibmq\_montreal).\cite{Pereira2022} They tested a set of four states, three of them compatible with those in our work: two separable states and Bell/GHZ states (in addition, they tested a product of Bell states). For separable states, the experiment was performed for up to ten qubits with measurements in (i) separable bases, and up to four qubits with measurements in (ii) nonseparable bases; for the Bell/GHZ state, the experiment was carried out for up to four qubits in both cases. Comparing the results for equivalent states and number of qubits, we have that for separable states the PZD method (i) obtained fidelities above $0.99$ for all cases from $n=2$ to 5, and were above our best results, whereas the method (ii) presented equivalent fidelities for two and three qubits, and worse for four qubits. On the other hand, for the Bell/GHZ states, our best results were, in general, better than those of both PZD methods, as shown in Table~\ref{tab:PZDxPtycho}.

\begin{table}[t]
\centering
\caption{Experimental estimation fidelities of the Bell/GHZ states for $n=2,3,4$ qubits obtained by the PZD method (using both separable and nonseparable bases) and the ptychographic method (using both QFT and AQFT of degree $2$).}
\label{tab:PZDxPtycho}
\begin{ruledtabular}
\begin{tabular}{lcccc}
& \multicolumn{2}{c}{PZD method \cite{Pereira2022}}  & \multicolumn{2}{c}{Quantum ptychography}  \\[1mm] 
& Separable & Nonseparable & QFT & AQFT ($m=2$) \\[0.5mm]
\hline \\[-2.5mm]
\multicolumn{1}{l}{$|\psi_+\rangle$} & $0.972$ & $0.971$ & $0.985$ & ---  \\\multicolumn{1}{l}{$|\textrm{GHZ}\rangle_3$} & $0.902$ & $0.904$ & $0.912$ & $0.943$  \\ 
\multicolumn{1}{l}{$|\textrm{GHZ}\rangle_4$} & $0.864$ & $0.693$ & $0.672$ & $0.894$ \\
\end{tabular}
\end{ruledtabular}
\end{table}

This comparison shows interesting aspects of both methods, but it must be analyzed bearing in mind that the experiments were carried out both in different periods and on different IBM processors. The \textsc{cnot} gates on ibmq\_manhatan were reported\cite{Pereira2022} to have an average error of $2\times 10^{-2}$, which is $2.5$ higher than in the processor used here (see Table~\ref{tab:ibm_devices}). Moreover, the maximum allowed number of shots at this device was $2^{13}$, while in ours, it was $10^5$. Despite the more favorable conditions, the results of the ptychographic method we are discussing here were obtained from measurements in nonseparable bases only (whether using QFT or AQFT), and in many cases, they were comparable or even better than those from PZD. In addition, unlike Ref.~\onlinecite{Pereira2022}, we have also performed the estimation of the $n$-qubit $W$ state, which requires the most complex circuit for its generation \cite{Cruz2019} among all tested states in both works and consequently is the most affected by errors in its preparation. For $n=3$ and $4$ qubits, we obtained fidelities $F'_W=0.949$ and $F'_W=0.879$, respectively, with measurements in the nonseparable basis generated by the AQFT with $m=2$, showing the robustness of the ptychographic method.

\subsection{QFT vs AQFT}

The QFT is a key operation for many quantum algorithms.\cite{NielsenBook} In particular, concerning its application in the period-finding problem, Barenco \textit{et al.}\cite{Barenco1996} showed that in a noise-free scenario, the performance of an AQFT of degree $m>\log_2n+2$ is almost identical to the exact QFT itself. They further showed that under noise, the AQFT yields better results than the QFT, with the optimal $m$ approaching the lower bound. Here, we observed a similar behavior in the ptychographic protocol regarding the use of QFT and AQFT. Although our limited set of results prevents us from drawing conclusions about this behavior, we can still conjecture the underlying causes.  The two conjectures presented next are based on the protocol carried out with the projectors of Eq.~(\ref{eq:Pi_xiq}). First, like in the period-finding problem, there is a lower bound for $m$ in the absence of noise that maintains the accuracy of the ptychographic estimation for a given $n$. If so, this would limit the degrees of approximation useful in a noisy scenario. Second, any nonseparable measurement basis unbiased with respect to the computational one might be effective within the protocol. If so, any degree of approximation $m>1$ would work for quantum ptychography. The results of the noise-free simulations shown in Fig.~\ref{fig:AQFTsimulations} seem to indicate this trend but the sampling is too small to be conclusive. These conjectures should guide future investigations, and regardless of the conclusion reached, there is still an immediate limitation to using the AQFT of degree $m>1$ due to its nonseparable nature, which remains unfavorable in current quantum processors.

\subsection{Further steps toward scalability on noisy devices}  \label{subsec:FurtherSteps}

The noise level on quantum devices like the one tested in this work severely affects the ptychographic method that combines separable intermediate projections with a final measurement in the nonseparable basis generated by the inverse QFT. In this scenario, the final measurement in a nonseparable basis generated by a unitary operation with a less noisy circuit, such as the AQFT, has shown to be a promising path toward the scalability of the method. On the other hand, implementing the final measurement on a separable basis would be the best option to reduce noise effects. However, we have seen that while this approach worked very well for separable states, it will not work in general for entangled states. A possible way to circumvent this issue is to combine the final separable basis with nonseparable intermediate projections in small subspaces of the $n$-qubit system. For instance, one could implement projections onto Bell states between connected physical qubits, which would generate less noisy ptychographic circuits than those that use QFT or even AQFT. This will be the subject of future investigations.

Another intervention in the ptychographic method to deal with noise in quantum devices is to improve the postprocessing algorithm itself. As mentioned earlier, here used its simplest form. Nevertheless, we have seen that a small modification with the introduction of a variable feedback parameter has already produced significant improvements in the estimation of the states. More sophisticated versions of the PIE algorithm have been introduced recently and shown to be much more robust to handle noisy datasets.\cite{Maiden2017}

In addition to interventions at both measurement and postprocessing stages of the protocol, error mitigation could also be a useful tool for its scalability. Here, we have applied a measurement error mitigation technique based on a very simple model. Despite its low effectiveness, especially for larger $n$, we witnessed improvements in most estimation fidelities. Addressing error mitigation through more sophisticated and robust techniques\cite{Bravyi2021,Nation2021,Kim2023} will certainly have a stronger impact on the estimations.

\section{Conclusion}
\label{sec:Conclusion}

We have presented a comprehensive study of quantum ptychography applied for estimating $n$-qubit pure states in a circuit-based quantum computer. According to the original proposal,\cite{Fernandes2019} we considered the protocol implemented through separable overlapping projections, given by Pauli measurements on one qubit at a time, followed by a projective measurement in a nonseparable basis generated by the inverse QFT. The number of circuits, in this case, scales linearly with $n$, which is comparable to the PZD method\cite{Pereira2022} and advantageous over standard tomographic schemes, where the scaling is exponential.\cite{Altepeter2005} In this scenario, we first implemented numerical simulations for up to $n=10$ qubits using fixed ensemble sizes; it was shown that the ptychographic method outperformed PZD for large $n$. The experimental implementation on an IBM quantum processor for up to five qubits achieved high estimation fidelities for separable and entangled states of $n=2$ and $n=3$ qubits, only. 

For $n>3$ qubits, the decline of the fidelities was mainly due to the high noise from the QFT circuit. To circumvent this, we investigated alternative unitary operations for ptychography with less noisy circuits. First, we used the AQFT and showed its potential to replace the QFT through numerical simulations. The experimental implementation using the AQFT of degree $m=2$, led us to significant improvements in the estimation fidelities in all cases, despite the low fidelities for the $5$-qubit entangled states caused mostly by their preparation circuits. Second, we used separable unitary operations for which we obtained high estimation fidelities only for separable states of up to $n=7$ qubits; in this case, the difficulty with entangled states was more related to the separable nature of the measurement basis than to noise effects. 

Other aspects that we can highlight in this work are the following: first, our experimental results, especially for the estimation of entangled states, were also more favorable compared to those of PZD; second, in our experiments, we applied a very simple measurement error mitigation technique that, although not very effective, improved almost all estimation fidelities; finally, we implemented a modification in the postprocessing algorithm which greatly improved its convergence. 

In addition to demonstrating quantum ptychography on a quantum computer and implementing modifications to the method aiming at its scalability in noisy devices, our study also opens up some possibilities to investigate further improvements. For instance, the study of new families of intermediate projections combined with a separable basis for the final measurement, which will reduce noise in the ptychographic circuits, as well as the application of more advanced error mitigation techniques. This will be essential in the short term to ensure scalability, so that in the future, with less noisy quantum processors, the ptychographic method can become a valuable tool for assessing the performance of such devices.

\begin{acknowledgments}
This work was supported by CNPq INCT-IQ (465469/2014-0), CNPq (422300/2021-7) and (303212/2022-5). We also acknowledge the Brazilian agencies CAPES and FAPEMIG. W.M.S.A. acknowledges financial support from CNPq (130062/2021-9). We acknowledge the use of IBM Quantum services for this work. The views expressed are those of the authors, and do not reflect the official policy or position of IBM or the IBM Quantum team.
\end{acknowledgments}

\section*{Author declarations}
\subsection*{Conflict of Interest}
The authors have no conflicts to disclose.

\subsection*{Author Contributions}
\noindent \textbf{Warley M. S. Alves:} Formal analysis (equal); Investigation (lead); Methodology (equal); Software (lead); Writing –
review \& editing (equal). \textbf{Leonardo Neves:} Conceptualization (lead); Formal analysis (equal); Investigation
(supporting); Methodology (equal); Supervision (lead); Writing – original draft (lead); Writing – review \& editing
(equal).

\section*{Data Availability}
The data that support the findings of this study are available from the corresponding author upon reasonable request.

\section*{References}
\vskip-0.5cm
\nocite{*}
\input{PtyNISQ.bbl}

\end{document}

%% file: PtyNISQ.bbl
%